\RequirePackage{rotating}
\documentclass[iop]{emulateapj} 
\usepackage{xspace}
\usepackage{natbib}
\usepackage{hhline}
\usepackage{amsmath}
\usepackage{dcolumn}
\usepackage{verbatim}
\usepackage{calc}
\usepackage{color}
\usepackage{booktabs}
\usepackage{longtable}
\usepackage[bookmarks]{hyperref}
\usepackage[numbered]{bookmark}
\usepackage{rotating}
\usepackage{paralist}
\usepackage[multidot]{grffile}
\usepackage{tabto}
\usepackage[usenames,dvipsnames,svgnames,table]{xcolor}

\newcolumntype{.}{D{.}{.}{-1}}

\newcommand{\chandra}{{\it Chandra}\xspace}

\newcommand{\xmm}{{\it XMM-Newton}\xspace}
\newcommand{\rxte}{{\it RXTE}\xspace}

\newcommand{\sS}[1]{\mbox{$\rm{}^{#1}$}}
\newcommand{\Ss}[1]{\mbox{$\rm{}_{#1}$}}
\newcommand{\nep}[2]{\mbox{${#1}$$\times$${10}^{#2}$}}

\newcommand{\Ms}{\mbox{$M_{\odot}$}\xspace}
\newcommand{\Zs}{\mbox{$Z_{\odot}$}\xspace}
\newcommand{\Rs}{\mbox{$R_{\odot}$}\xspace}
\newcommand{\nH}{\mbox{$N$\Ss{H}}\xspace}

\newcommand{\Deg}{\mbox{$^\circ$}\xspace}
\newcommand{\x}{\mbox{$\times$}}

\newcommand{\lcgs}{\mbox{erg s\sS{-1}}\xspace}
\newcommand{\fcgs}{\mbox{erg s\sS{-1} cm\sS{-2}}\xspace}

\definecolor{ao}{rgb}{0.0, 0.5, 0.0}
\definecolor{amber}{rgb}{1.0, 0.49, 0.0}
\definecolor{ballblue}{rgb}{0.13, 0.67, 0.8}
\definecolor{bleudefrance}{rgb}{0.19, 0.55, 0.91}
\definecolor{brandeisblue}{rgb}{0.0, 0.44, 1.0}

\begin{document}

\title{
SXP214, an X-ray Pulsar in the Small Magellanic Cloud,
Crossing the Circumstellar Disk of the Companion
}



\author{
JaeSub~Hong\altaffilmark{1},
Vallia~Antoniou\altaffilmark{1},
Andreas~Zezas\altaffilmark{1,2,3},
Frank~Haberl\altaffilmark{4},
Jeremy~J.~Drake\altaffilmark{1},
Paul~P.~Plucinsky\altaffilmark{1},
Terrance~Gaetz\altaffilmark{1},
Manami~Sasaki\altaffilmark{5},
Benjamin~Williams\altaffilmark{6},
Knox~S.~Long\altaffilmark{7},
William~P.~Blair\altaffilmark{8},
P.~Frank~Winkler\altaffilmark{9},
Nicholas~J.~Wright\altaffilmark{10},
Silas~Laycock\altaffilmark{11},
Andrzej~Udalski\altaffilmark{12} 
}
\altaffiltext{1}{Harvard-Smithsonian Center for Astrophysics, 60 Garden St., Cambridge, MA 02138, USA: jaesub@head.cfa.harvard.edu} 
\altaffiltext{2}{Foundation for Research and Technology-Hellas, 71110 Heraklion, Crete, Greece}
\altaffiltext{3}{Physics Department \& Institute of Theoretical \& Computational Physics, University of Crete, 71003 Heraklion, Crete, Greece}  
\altaffiltext{4}{Max-Planck-Institut f{\"u}r extraterrestrische Physik, Giessenbach stra\ss e, 85748 Garching, Germany} 
\altaffiltext{5}{Institut f{\"u}r Astronomie und Astrophysik, Universit{\"a}t T{\"u}bingen, Sand 1, 72076, T{\"u}bingen, Germany}  
\altaffiltext{6}{Department of Astronomy, Box 351580, University of Washington, Seattle, WA 98195, USA} 
\altaffiltext{7}{Space Telescope Science Institute, 3700 San Martin Drive, Baltimore, MD 21218, USA} 
\altaffiltext{8}{The Henry A. Rowland Department of Physics and Astronomy, Johns Hopkins University, 3400 North Charles Street, Baltimore, MD 21218, USA} 
\altaffiltext{9}{Department of Physics, Middlebury College, Middlebury, VT 05753, USA} 
\altaffiltext{10}{Astrophysics Group, Keele University, Keele, ST5 5BG, UK}
\altaffiltext{11}{Department of Physics, University of Massachusetts Lowell, MA 01854, USA} 
\altaffiltext{12}{Warsaw University Observatory, Al. Ujazdowskie 4, 00-478 Warszawa, Poland}

\begin{abstract}

Located in the Small Magellanic Cloud (SMC), SXP214 is an X-ray pulsar
in a high mass X-ray binary system with a Be-star companion.  A recent
survey of the SMC under
a \chandra X-ray Visionary program found the source in a transition
when the X-ray flux was on a steady rise.  The Lomb-Scargle periodogram
revealed a pulse period of 211.49 $\pm$ 0.42 s, which is significantly
($>$5$\sigma$) shorter than the previous measurements with
\xmm and \rxte. This implies that the system has gone
through sudden spin-up episodes recently.
The pulse profile shows a sharp eclipse-like feature with a modulation
amplitude of $>$95\%.  The linear rise of the observed X-ray
luminosity from $\lesssim$2\x\ to \nep{7}{35} \lcgs is correlated
with steady softening of the X-ray spectrum, which can be described by
the changes in the local absorption from \nH $\sim$ 10\sS{24}
to $\lesssim$ 10\sS{20}
cm\sS{-2} for an absorbed power-law model.  The soft X-ray emission
below 2 keV was absent in the early part of the observation when only
the pulsating hard X-ray component was observed, whereas at later
times both soft and hard X-ray components were observed pulsating.
A likely explanation is that the neutron star was initially hidden in the
circumstellar disk of the companion, and later came out of the disk
with the accreted material that continued fueling the observed
pulsation.


\end{abstract}

\keywords{stars: neutron --- X-rays: binaries}

\section{Introduction} \label{s:intro}

The Small Magellanic Cloud (SMC) harbors a large number of high mass
X-ray binaries with pulsating neutron stars (NS) and Be-star companions
\citep{Coe15b, Haberl15}.  In a
system with a relatively long orbital period, the NS is often in a wide
and eccentric orbit, and periodically encounters the circumstellar
disk of the Be star, which triggers Type-I outbursts through accretion onto
the NS \citep{Reig11, Townsend11}.  SXP214 or XMMU J005011.2--730026 was
discovered as a transient pulsar in the SMC with a pulse period of $P$
= 214~s from an \xmm observation in 2009 
\citep[][hereafter C11]{Coe11}.  A B2-B3 III star with $V$~=~15.3 mag
and a weak H$\alpha$ emission line
was identified as the optical counterpart from the OGLE-II \& III
data and an optical spectrum taken at the South African Astronomical
Observatory (C11).  Thus, SXP214 has been classified as a Be/X-ray
binary (Be-XRB; C11).

The orbital period of the system remains unclear. C11 reported an optical
periodicity of 4.52~d based on the OGLE-III lightcurve.
\citet{Schmidtke11} reported another optical period of 0.82~d based on
the MACHO and OGLE-II observations, arguing that the 4.52~d periodicity is
a weaker alias of the 0.82~d period and the latter is more consistent
with other Be-XRBs.  Later \citet{Schmidtke13} reported 
a new optical period of 29.91~d with an eclipse-like
feature in the folded lightcurve from the OGLE-IV data
when the optical counterpart became faint and the 0.82~d optical
pulsations disappeared.

Here we describe new \chandra observations of SXP214 (\S\ref{s:obs}),
which show dramatic temporal and spectral changes of the X-ray emission
(\S\ref{s:overview}).  We search for periodicity in the X-ray
emission and perform phase-resolved spectral analysis
to understand the origin of the observed changes
(\S\ref{s:timing}). We revisit the OGLE optical lightcurve and its
long term periodicity (\S\ref{s:opt}). We present a likely scenario for
the changes observed in the X-ray emission and constrain the density
of the circumstellar disk, the magnetic field of the NS and the
accretion rate (\S\ref{s:discussion}).

\begin{figure*} \begin{center}
\includegraphics*[width=0.950\textwidth,clip=true] {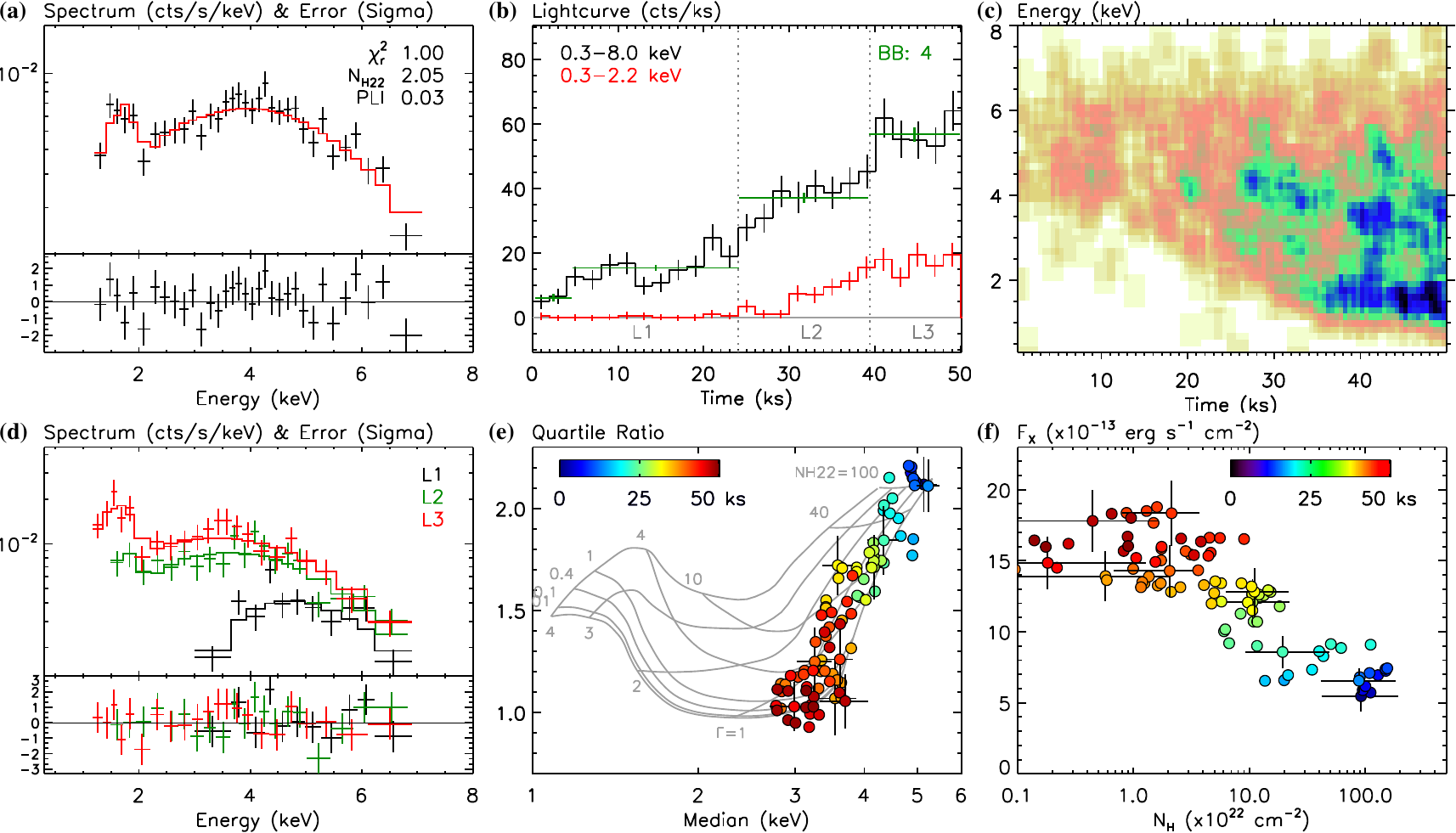}
\caption{
(a) The \chandra spectrum and the best-fit absorbed power-law model. 
(b) The \chandra 0.3--8 keV (black) and 0.3--2.2 keV (red) lightcurves
overlaid with the Bayesian blocks (green). 
(c) The temporal energy distribution of the X-ray events in the source
aperture region. Dark shades or blue colors indicate high counts and 
light shades or yellow colors indicate low counts. 
(d) The absorbed power-law model fits for three intervals,
L1: the first 24 ks, L2: the middle 15 ks, and L3: the last 11 ks (Table~\ref{t:spec}). 
(e) The time-segmented quantile diagram (25\% to 75\% quartile ratios vs.~median energy). 
The grids are for an absorbed power-law model
covering photon indices of $\Gamma$ = 0, 1, 2, 3, and 4, and local
absorptions of \nH = 0.01, 0.1, 0.4, 1, 4, 10, 40, and 100 $\times$ 10\sS{22} cm\sS{-2}.
The errors of several data points are shown for illustration.
(f) The observed 0.5--8 keV X-ray flux vs.~the quartile-ratio based estimate of
the local absorption under the assumption of $\Gamma$=0.5.  
The symbols in (e) and (f) are color-coded with the elapsed observing time.
}
\label{f:g}
\end{center}
\end{figure*}

\section{Observations and Data Processing} \label{s:obs}

SXP214 was observed in 2013 January (Obs.~ID 14670) and 2014 March
(Obs.~ID 15503) by \chandra for 50
ks each as a part of the recent SMC survey under the \chandra X-ray
Visionary Program (PI A.~Zezas). The details of the survey
program and the full source catalog are found in \citet{Antoniou16}. 
The X-ray timing analysis and the catalog of the SMC pulsars in the
survey are discussed by \citet{Hong16}.
The data presented here were processed and analyzed by the latest X-ray
analysis pipeline using CIAO ver 4.6 
developed for the \chandra Multi-wavelength Plane survey
\citep[][]{Grindlay05, Hong12}. 

The source was observed $\sim$7.5\arcmin\  off the aimpoint
in both observations. The source was detected in the first observation
with 1511 net counts in the 0.3--8 keV band
or a 0.5--8 keV X-ray luminosity of \nep{4.7}{35} \lcgs at 60 kpc (see
below for the spectral model used), while it was not 
detected in the the second observation
with an upper limit of 12 net counts at
3$\sigma$ significance ($L_X$ $\lesssim$ \nep{4}{33} \lcgs).
According to the \chandra archival data, a \chandra observation
in 2002 October (Obs.~ID 2945) also covered SXP214
at $\sim$17\arcmin\ off the aimpoint for 12 ks, but it was not detected
($L_X$ $\lesssim$ \nep{3}{34} \lcgs).
The following analysis is
based on the 2013 January observation (Obs.~ID 14670).
The \chandra position of the source is
R.A.~=~00\sS{h}50\sS{m}11.26\sS{s} and Decl.~=~$-$73{\Deg}00$'$25.6$''$
(J2000) with a 95\% error radius of 0.47\arcsec\, where the error 
is based on the formula by \citet{Hong05}. 

\section{Spectral and Temporal Evolution} \label{s:overview}

\begin{figure*} \begin{center}
\includegraphics*[width=0.950\textwidth,clip=true] {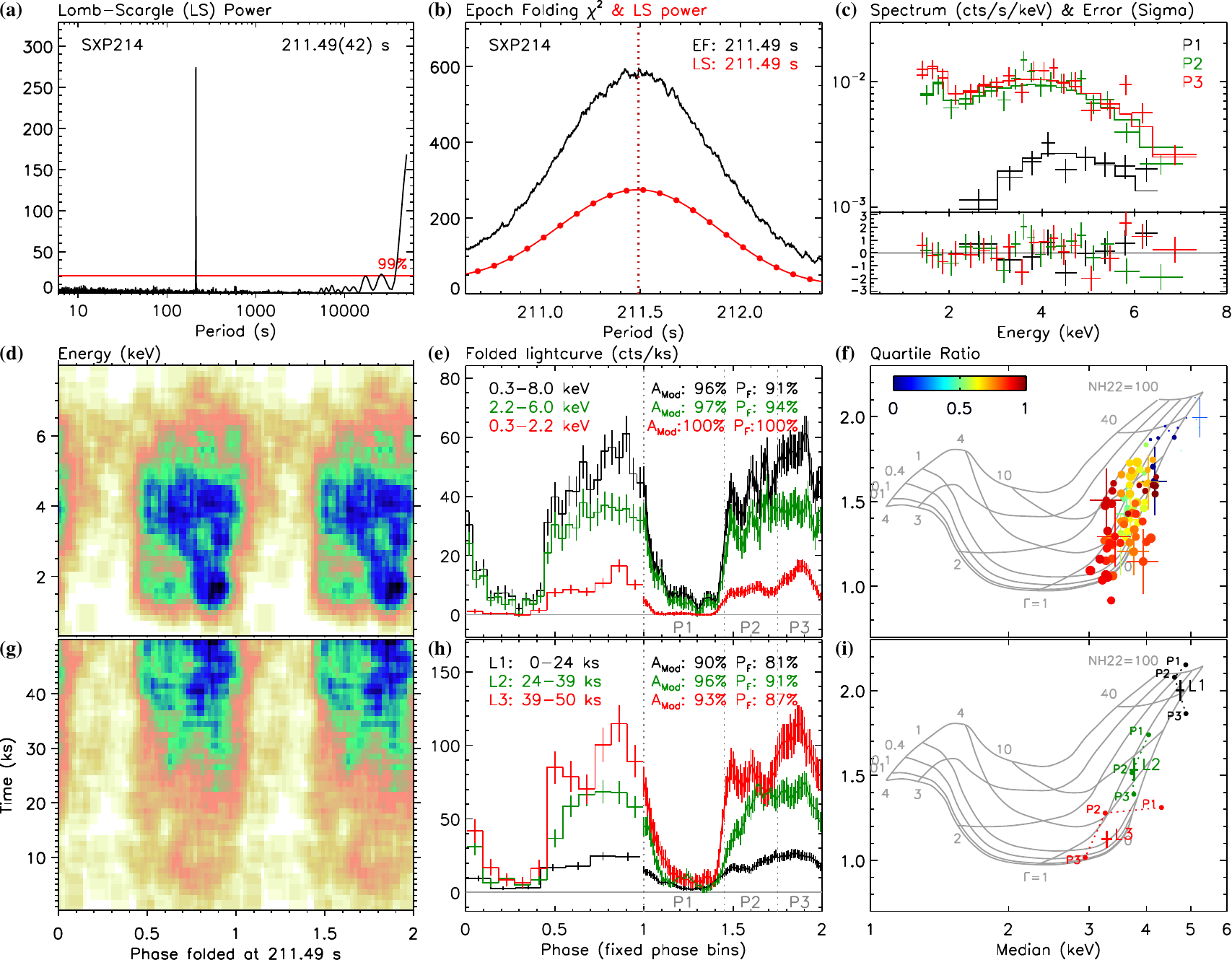}
\caption{
(a) The Lomb-Scargle periodogram with the 99\% confidence level.
(b) A close-up of the epoch folding and Lomb-Scargle periodograms
around the pulse period.
(c) The absorbed power-law model fits for three phase segments,
P1: 0.00 -- 0.45, P2: 0.45 -- 0.75, and P3: 0.75 -- 1.00.
(d) The energy vs.~pulse phase distribution of the X-ray events.
(e) The folded lightcurves in the 0.3--2.2 keV (red), 2.2--6.0 keV (green) and
0.3--8.0 keV (black) bands using fixed-size phase bins (phases 0--1)
and sliding window bins (phases 1--2).
(f) The phase-resolved quantile diagram. The symbol colors 
match the pulse phase, and the symbol sizes are
proportional to the count rates.
(g) The time vs.~pulse phase distribution of the X-ray events.
(h) The 0.3--8 keV folded lightcurves of the three time intervals, L1, L2, and L3.
(i) The quantile diagram for the three phase segments of each interval.
}
\label{f:p}
\end{center}
\end{figure*}

Fig.~\ref{f:g}{\it a} shows the overall X-ray spectrum and the best spectral
fit using an absorbed power-law model with two absorption components:
the Galactic foreground absorption fixed at \nH =
\nep{6}{20}~cm\sS{-2} \citep{Dickey90} with solar abundances ($Z$~=~$\Zs$), and the free
SMC and local absorption with reduced metal abundances of $Z$~=~0.2~$\Zs$
\citep{Russell92} following the absorption model by \citet{Wilms00}.
Each spectral bin is set to contain at least 40 net counts.
The best-fit photon index is $\Gamma$ = 0.0 $\pm$ 0.1 with \nH =
\nep{2.1}{22}~cm\sS{-2} and reduced $\chi^2$ = 1
(Table~\ref{t:spec}).  An absorbed blackbody, thermal bremsstrahlung
or APEC model fits the X-ray spectrum poorly (reduced $\chi^2$ = 1.4 -- 4).

The X-ray lightcurve of the source 
(Fig.~\ref{f:g}{\it b}) shows a steady rise with an average count rate
of 30 counts per ks.  Bayesian block analysis, 
that searches for the change points between time intervals of
statistically different rates \citep{Scargle13},
identified four separate blocks (green).  
Fig.~\ref{f:g}{\it c} visualizes a remarkable change in the spectrum
using the temporal energy distribution of the X-ray events in the source
aperture region.  The early part of the X-ray spectrum is limited to
hard X-ray photons above $\sim$3 keV, and the softer X-ray photons below
2 keV appeared only after $\sim$20 ks into the observation.

To quantify the spectral evolution, we divide the data set into three
time intervals according to the Bayesian blocks:  L1,
L2, and L3 for the first 24 ks (combining the first two blocks with relatively low
counts), the middle 15 ks, and the last 11 ks, respectively.
Fig.~\ref{f:g}{\it d} shows the best-fit joint spectral model 
of the three intervals with a common photon index.
Each spectral bin of the segmented data set is set to contain at least
20--30 net counts.
Table~\ref{t:spec} lists the best-fit parameters for both
joint and individual fits, which are all consistent with $\Gamma$ = 0.5,
indicating that the spectral changes are driven by the
variation in the local absorption.

Fig.~\ref{f:g}{\it e} illustrates the evolution of spectral parameters 
in a finer timescale using the energy quantile diagram 
(quartile ratio vs.~median energy), which can classify
diverse spectral types without spectral bias often inherent to X-ray hardness
or X-ray color-color diagrams \citep{Hong04}.
Each data point contains about 75 net counts in sliding time windows.
The grids are for an absorbed power-law model covering a wide range of
photon indices ($\Gamma$~=~0--4) and local absorptions (\nH =
10\sS{20--24} cm\sS{-2}).
The rise in the count rate from $\sim$ 5 to 60 cts ks\sS{-1} is well
correlated with the significant reduction in the local absorption from
\nH $\sim$ 10\sS{24} to 10\sS{20} cm\sS{-2} for the
aforementioned absorbed power-law model, whereas the photon index
remains more or less constant. Fig.~\ref{f:g}{\it f} shows the
correlation between the observed 0.5--8 keV X-ray flux and the local
absorption, assuming $\Gamma$~=~0.5, whereas the intrinsic X-ray flux or luminosity
remains more or less constant within the uncertainty (Table~\ref{t:spec}). 

\section{Timing Analysis} \label{s:timing}

\begin{figure*} \begin{center}
\includegraphics*[width=0.850\textwidth,clip=true] {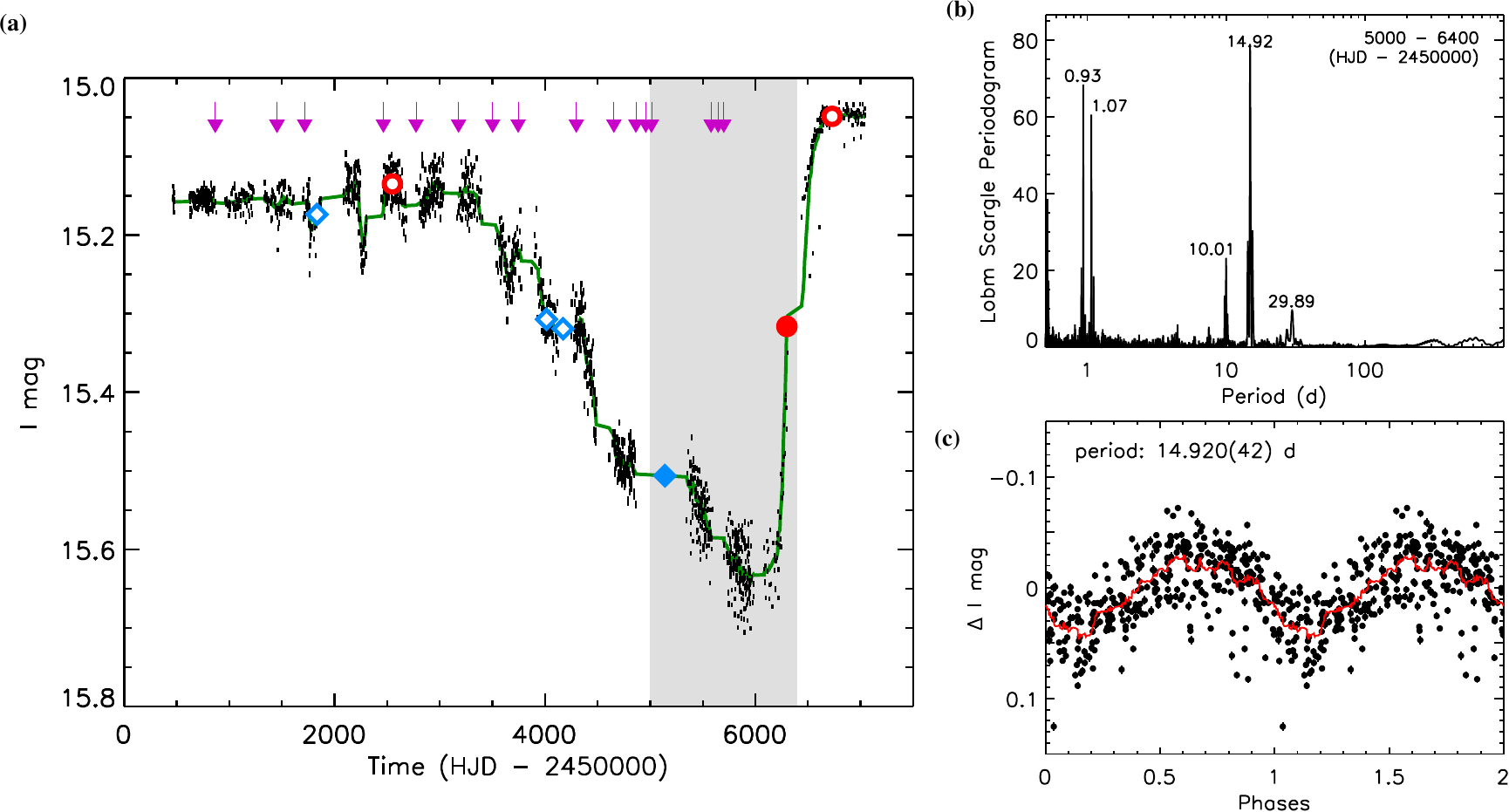}
\caption{(a) The OGLE-II, III \& IV lightcurve, 
(b) the Lomb-Scargle periodogram for the 1400 days starting from HJD
2455000 (the shaded region of the lightcurve), and 
(c) the folded lightcurve of the same data at a period of 14.920 d.
In (a), the green line represents the trend calculated
by smoothing the lightcurve with a 20 data window. The blue diamonds and red circles
represent the time of \xmm and \chandra observations, respectively,
and the solid symbols indicate the detection of the X-ray
periodicity. The pulsation detections by \rxte are marked by the magenta arrows.
}
\label{f:o}
\end{center}
\end{figure*}

We search for periodic X-ray modulations in the 0.3--8 keV lightcurve using a
Lomb-Scargle (LS) periodogram \citep{Scargle82}.  For the details of
the timing analysis, see  \citet{Hong16}. Fig.~\ref{f:p}{\it a} plots the LS periodogram over
the entire search range, revealing a sharp peak at $\sim$ 211.5 s. An
increase in the power of the periodogram at periods longer than 10 ks
is due to the long-term rise of the lightcurve.
We refine the modulation period with a follow-up search using the epoch
folding (EF) and LS methods around the initial period. Both EF and LS
periodograms peak at the same period of 211.49 $\pm$ 0.42~s (Fig.~\ref{f:p}{\it b}).  
The error of the modulation period is a 1$\sigma$-equivalent width (34\%) of
the peak in the periodogram, which tends to be a bit
conservative, relative to the error estimate (0.13 s) based on \citet{Horne86}.  
Nonetheless, the newly detected pulse period is not only significantly shorter 
than the previously reported period of 
214.045 $\pm$ 0.052 s by C11 from the \xmm observation in 2009 November,
but it is also against the long-term spin-down trend ($\dot{P}$ $\sim$ +0.12 s yr\sS{-1}
with the average period of 213.7 $\pm$ 0.1 s)
measured by \rxte \citep[][hereafter K14]{Klus14}. On the other hand,
the individual period measurements by K14 show a large scatter ranging
from 211 to 216 s, and two out of 17 \rxte measurements show the
periods consistent with our result.

Fig.~\ref{f:p}{\it d} illustrates
the drastic spectral variation with pulse phase using
the energy vs.~phase distribution of the X-ray events.
The energy-band selected pulse profiles in Fig.~\ref{f:p}{\it e} 
show large peak-to-valley amplitudes ($A\Ss{mod}$ $\ge$
95\%) and high pulse fractions ($P_F$ $\ge$ 91\%),\footnote{
$A\Ss{mod}$ is defined as
1--$r$\Ss{min}/$r$\Ss{max} where $r$\Ss{min} and $r$\Ss{max} are the
minimum and maximum of the folded lightcurve. $P_F$ is defined
as the ratio of the pulsating flux above the minimum to the total
flux. i.e., $\Sigma_i (r_i-r\Ss{min})/ \Sigma_i r_i$, where $r_i$ is the rate of
the folded bin $i$.} which contrasts with a
somewhat marginal $P_F$ of 29\% measured by \xmm (C11). 
The \chandra pulse profile shows clear on- and
off-phases of almost equal durations. 
The soft X-ray emission below 2 keV appears fully
`eclipsed' during the off-phase and exhibits a two-step rise during
the on-phase, whereas the hard X-ray emission is more or less uniform
during the on-phase with some trailing into the off-phase.  

We divide the pulse phases into three segments based on the
soft-band pulse profile:
P1 for the off-phase covering 0.00 -- 0.45, P2 for the first part of
the rise covering 0.45 -- 0.75, and P3 for the peak covering 0.75 -- 1.00.
Fig.~\ref{f:p}{\it c} shows the best-fit joint model of the three
segments with a common photon index and Table~\ref{t:spec} lists the
best-fit parameters for both joint and individual fits.  When $\Gamma$ $\lesssim$
0 (P1), the photon index becomes somewhat degenerate with the absorption,
so the best-fit parameters can be misleading.  On the other hand, 
in the joint fit and the phase-resolved quantile diagram
in Fig.~\ref{f:p}{\it f} there {\it appears} to be a clear trend that
the spectral evolution over pulse
cycles is also dominated by the changes in the local absorption.
At the peak of the pulse profile (phase $\sim$ 0.9), the local
absorption drops below \nH = 10\sS{20} cm\sS{-2}, whereas during the
off-phases ($\sim$ 0.1 to 0.5) it rises well above 10\sS{23} cm\sS{-2}.
The quantile diagram in Fig.~\ref{f:p}{\it i} compares the spectral types
of the three phases of each time interval. The spectral evolution over
the pulse phase within each interval is not dramatic, indicating
that the spectral evolution seen in Figs.~\ref{f:p}{\it c} and
\ref{f:p}{\it f} is likely amplified by the temporal evolution.


Fig.~\ref{f:p}{\it g} shows the time vs.~pulse phase distribution
of the X-ray events. 
The pulsation is present from the very beginning
of the observation when no soft X-ray emission below 2 keV was observed.
The apparent skewness in the distribution (e.g, at phase $\sim$ 0),
which is usually an indication of an inaccurate period estimate
\citep[][]{Hong16}, is an artifact due to
the flux change with the time: i.e., the distribution,
if renormalized at each time bin, would not be skewed (not shown).
Fig.~\ref{f:p}{\it h} shows the
folded lightcurve of each time interval, and the strong pulsation is
observed in all three intervals ($A$\Ss{mod} $\ge$ 90\%).


\section{OGLE Lightcurve Analysis} \label{s:opt}

Fig.~\ref{f:o}{\it a} shows the optical lightcurve in $I$ mag spanning over
20 years using the data from the OGLE-II, III, and IV surveys. The red
circles and blue diamonds mark 
the three \chandra and four \xmm observations, respectively. 
The X-ray pulsations (solid symbols) were detected when the optical
lightcurve was experiencing a significant drop or recovering from
the drop. \rxte also detected the X-ray pulsations from the source
multiple times (purple arrows; K14).

Given the uncertainty in the orbital period of the system, we search
for periodic modulation in the full OGLE lightcurve using the LS
periodogram.  The LS periodogram of the raw lightcurve shows three
dominant periods of 0.5, 1, and 348~d. Detrending techniques eliminate
these false periods to varying degrees but the final periodicity
results are not sensitive to the detrending techniques.  We use the
moving average of 20 data points as the lightcurve trend. 

We also calculate the LS periodogram of the OGLE lightcurve using sliding
windows of 500~d to identify any changes in the periodicity.  Indeed the
periodicity transitions in about MJD 55000, and then the last 500~d of the
lightcurve do not show any clear sign of periodicity.  In the first 5000
days of the lightcurve 0.82 and 4.58~d periods are most dominant as
reported by C11 and \citet{Schmidtke11}.  We note that the 0.82~d period
is a beat period among 0.5, 1, and 4.58~d periods. Fig.~\ref{f:o}{\it
b} shows the periodogram of the next 1400 days where
a 14.92~d period is most prominent. Its second harmonic (29.89~d),
which is consistent with the 29.91~d period reported by
\citet{Schmidtke13}, is marginally significant. The 0.93 and
1.07~d periods are the beating periods among 0.5, 1,
and 14.92~d.  Fig.~\ref{f:o}{\it c} shows the folded lightcurve at
the 14.92~d period.  The analysis of the full OGLE lightcurve did not
reveal any other new periods.
Given their instability, it appears premature to claim that 
any of the above periods is the orbital period

\begin{table*}
\begin{minipage}{0.99\textwidth}
\caption{Spectral analysis using an absorbed power-law model}
\begin{tabular*}{\textwidth}{r r@{\ }l r@{}l r@{}l r@{}l  r@{\ }l r@{}l r@{}l r@{}l r@{}l}
\hline\hline
   \multicolumn{1}{c}{(1)}                                     & \multicolumn{2}{c}{(2)       } & \multicolumn{2}{c}{(3)     } & \multicolumn{2}{c}{(4)                                    } & \multicolumn{2}{c}{(5)         } & \multicolumn{2}{c}{(6)             } & \multicolumn{2}{c}{(7)                          } & \multicolumn{2}{c}{(8)      } & \multicolumn{2}{c}{(9)                         } & \multicolumn{2}{c}{(10)     } \\
Data Segment                                                   & \multicolumn{2}{c}{Net Count } & \multicolumn{2}{c}{$\Gamma$} & \multicolumn{4}{c}{\nH in 10\sS{22} cm\sS{-2}             } & \multicolumn{2}{c}{$\chi^2_r$ / DoF} & \multicolumn{4}{c}{$F$\Ss{X}} & \multicolumn{4}{c}{$L$\Ss{X}} \\
\cmidrule(r){6-9}\cmidrule(r){12-15}\cmidrule(r){16-19}        & \multicolumn{2}{c}{0.3--8 keV} & \multicolumn{2}{c}{        } & \multicolumn{2}{c}{joint fit                              } & \multicolumn{2}{c}{(indiv.~fit)} & \multicolumn{2}{c}{                } & \multicolumn{2}{c}{Observed                     } & \multicolumn{2}{c}{Intrinsic} & \multicolumn{2}{c}{Observed                    } & \multicolumn{2}{c}{Intrinsic} \\
\hline
\multicolumn{1}{l}{\it All (Fig.~1{\it a})}                    & 1511       &  $\pm$ 39 & $0.0$    & $_{-0.1}^{+0.1}$  & \multicolumn{4}{c}{2.1$_{-0.7}^{+0.7}$} & 1.00             & / 31  & 10.9                          & $_{-1.7}^{+2.1}$ & 11.6      & $_{-1.8}^{+2.2}$ & 4.7                          & $_{-0.7}^{+0.9}$ & 5.0       & $_{-0.8}^{+0.9}$ \\
\hline\multicolumn{3}{l}{\it By elapsed time (Fig.~1{\it d})}  & 0.5      & $_{-0.1}^{+0.1}$  &                                         &                    &              &                     & 0.90             & / 46  &                               &                  &           &                  &                              &                  &           &                  \\
0 -- 24 ks (L1)                                                & 326        &  $\pm$ 18 & ($1.2$   & $_{-1.0}^{+1.1}$) & 70.3                                    & $_{-10.9}^{+11.6}$ & (91.9        & $_{-35.3}^{+38.3}$) & (1.43            & / 8)  & 7.0                           & $_{-1.8}^{+2.4}$ & 14.0      & $_{-3.5}^{+4.7}$ & 3.0                          & $_{-0.8}^{+1.0}$ & 6.0       & $_{-1.5}^{+2.0}$ \\
24 -- 39 ks (L2)                                               & 567        &  $\pm$ 24 & ($0.5$   & $_{-0.2}^{+0.3}$) & 4.9                                     & $_{-1.4}^{+1.4}$   & (5.0         & $_{-2.0}^{+2.1}$)   & (0.95            & / 17) & 12.7                          & $_{-3.2}^{+4.3}$ & 14.8      & $_{-3.7}^{+5.0}$ & 5.5                          & $_{-1.4}^{+1.8}$ & 6.4       & $_{-1.6}^{+2.1}$ \\
39 -- 50 ks (L3)                                               & 617        &  $\pm$ 25 & ($0.5$   & $_{-0.2}^{+0.2}$) & 1.2                                     & $_{-0.8}^{+0.8}$   & (1.1         & $_{-0.9}^{+0.9}$)   & (0.72            & / 19) & 16.0                          & $_{-4.0}^{+5.4}$ & 17.1      & $_{-4.3}^{+5.8}$ & 6.9                          & $_{-1.7}^{+2.3}$ & 7.4       & $_{-1.8}^{+2.5}$ \\
\hline\multicolumn{3}{l}{\it By folded phases (Fig.~2{\it c})} & 0.2      & $_{-0.1}^{+0.1}$  &                                         &                    &              &                     & 1.10             & / 40  &                               &                  &           &                  &                              &                  &           &                  \\
0.00 -- 0.45 (P1)                                              & 224        &  $\pm$ 15 & ($-1.3$  & $_{-0.3}^{+0.7}$) & 29.0                                    & $_{-7.8}^{+9.2}$   & (0.0         & $_{-0.0}^{+14.1}$)  & (0.59            & / 6)  & 4.5                           & $_{-1.0}^{+1.4}$ & 6.3       & $_{-1.5}^{+2.0}$ & 1.9                          & $_{-0.4}^{+0.6}$ & 2.7       & $_{-0.6}^{+0.8}$ \\
0.45 -- 0.75 (P2)                                              & 646        &  $\pm$ 25 & ($0.5$   & $_{-0.2}^{+0.2}$) & 3.2                                     & $_{-1.2}^{+1.2}$   & (5.0         & $_{-1.6}^{+1.7}$)   & (1.06            & / 16) & 14.8                          & $_{-3.5}^{+4.6}$ & 16.2      & $_{-3.8}^{+5.1}$ & 6.4                          & $_{-1.5}^{+2.0}$ & 7.0       & $_{-1.6}^{+2.2}$ \\
0.75 -- 1.00 (P3)                                              & 641        &  $\pm$ 25 & ($0.1$   & $_{-0.1}^{+0.2}$) & 1.4                                     & $_{-1.0}^{+1.0}$   & (0.4         & $_{-0.0}^{+1.3}$)   & (1.07            & / 16) & 16.3                          & $_{-3.8}^{+5.1}$ & 17.2      & $_{-4.0}^{+5.4}$ & 7.0                          & $_{-1.6}^{+2.2}$ & 7.4       & $_{-1.7}^{+2.3}$ \\
\hline
\end{tabular*}

\label{t:spec}
Notes-- 
(3) The best-fit photon index and
(4) \& (5) the SMC local absorption with 
the metal abundances of $Z$~=~0.2 $\Zs$ following \citet{Wilms00}
for an absorbed power-law model.
The fit parameters in the parentheses are from 
individual spectral fits instead of joint fits.
(6) The reduced $\chi^2$ and the degree of freedom (DoF).
(7) The absorbed and (8) unabsorbed 
0.5--8 keV energy fluxes in 10\sS{-13} \fcgs.
(9) The absorbed and (10) unabsorbed 0.5--8 keV luminosities at 60 kpc
 in 10\sS{35} \lcgs. Errors in the table represent 
the 68\% confidence (or $\pm$1$\sigma$ equivalent) interval.
\end{minipage}
\end{table*}

\section{Discussion} \label{s:discussion}

The typical orbital period of Be-XRBs with $\sim$200--300 s pulsations is
$\sim$30 to 200 days based on the so-called ``Corbet'' diagram
\citep[e.g.,][]{Reig07,Cheng14}.  K14 assumed 4.58~d as the orbital
period of SXP214, but according to their own orbital
vs.~spin period diagram, SXP214 is an outlier and 4.58~d appears too short
for the orbital period.
C11 detected weak H$\alpha$ emission (equivalent width:
$-$1.5$\pm$1.0 \AA) on MJD 55716, when the optical flux was near
its historical minimum. Since it is recognized that the continuum
variations in Be stars trace the size of the circumstellar disk \citep[e.g.,][]{Reig11},
the 14.92 or 29.89~d period, measured when the disk 
was relatively small, may represent the true
orbital period.  The observed historical maximum X-ray luminosity is
less than \nep{6}{36} \lcgs (K14).\footnote{K14 assumed $P_F$~=~33\% 
when converting the \rxte counts to the luminosity. Their
luminosity value can be an overestimate by a factor of three if the
pulsed fraction is close to 100\% as seen in this analysis. On the
other hand, K14 assumed $\Gamma$~=~1.5 for the
spectral model, which can lower their luminosity estimates by $\sim$30\% 
for a harder spectrum with $\Gamma$~=~0.5. Combining these two, the
luminosity estimate by K14 for SXP214 can be an overestimate by about
a factor of two.}
In fact, the system was detected only in two out of seven observations
(three with \chandra and four with \xmm), which is about 30\% duty cycle.
According to \citet{Reig07}, these
findings suggest that SXP214 is in a highly eccentric orbit ($\gtrsim$
0.3) with a long orbital period ($\gtrsim$ 20~d) and
should exhibit Type-I bursts ($<$10\sS{37} \lcgs).
Further they suggest that the circumstellar disk is likely only mildly
perturbed (but not fully truncated) by the NS. 

Our measurement of the spin period indicates that
the system may have gone through sudden spin-up episodes recently.
In a system with a highly elliptical orbit, accretion-induced spin up
episodes are likely to occur regularly, often near periastron passages.
The orbital speed ($v$) of the NS is 
\begin{eqnarray*}
	v  &=& \Big[ \frac{2 \pi G (M_1 + M_2)}{P\Ss{o}} \Big]^{1/3}
	\Big[ \frac{1+2 \epsilon \cos(\phi) + \epsilon^2}
	{1-\epsilon^2} \Big]^{1/2},
\end{eqnarray*}
where $G$, $M_1$, $M_2$, $\epsilon$, $\phi$, and $P_{\mbox{\scriptsize
o}}$ are the gravitational constant, the masses of the primary and
neutron stars, the eccentricity, the orbital phase, and the orbital
period, respectively. 
Within about 120\Deg of the periapsis (i.e., $|\phi|$$\lesssim$120\Deg)
which covers about 30\% of the orbital period,
for $\epsilon$ = 0.5, $M_1$ = 10 \Ms, and $M_2$ = 1.4 \Ms, 
$v$ $\sim$ 150 -- 270 ($M_{11.4}$/$P_{30}$)$^{1/3}$ km s\sS{-1} or  
150 -- 270 $a_{0.43}$ $P_{30}^{-1}$ km s\sS{-1},  
where $M_{11.4}$ is the combined mass in
units of 11.4 \Ms, $P_{30}$
is $P_{\mbox{\scriptsize o}}$ in units of 30 days,
and $a_{0.43}$ is the semi-major axis in units of 0.43 AU.  Therefore, the
apparent change in the spin period due to the orbital motion-induced
Doppler effect is less than $\sim$ 0.11 -- 0.19
($M_{11.4}$/$P_{30}$)$^{1/3}$ s for $\epsilon$ = 0.5, which is too
small to account for the observed period difference. 

According to \citet{Cheng14}, 
accretion through a disk can be an efficient spin-up process for the NS
whereas quasi-spherical or advection dominated accretion is relatively
inefficient.
Even in the case that the accretion is always through a disk, if the
accretion disk around
the pulsar is frequently reformed near periastron passages due to a
highly eccentric orbit, it is not surprising
to see the spin-up and spin-down torque reversal depending whether
each newly formed disk is prograde or retrograde relative to the NS spin
\citep{Nelson97}. Given the long-term spin-down trend even with
the two earlier \rxte measurements showing similarly short periods
(K14), the spin-up and spin-down torque reversal may occur routinely
in SXP214 in addition to the relatively slow spin-down trend due to
the rotational energy loss when there is no accretion. 

Roughly two years have passed between the last measurement of the spin
period by K14 and our measurements. Indeed the last few measurements
of the spin period by K14 already show a possible spin-up trend.
If we assume the recent spin-up
trend has lasted $\sim$ 1 -- 3 yrs, then $-\dot{P}$ $\sim$ 0.56 -- 1.7
s yr\sS{-1}.
For the accretion torque-induced spin-up model by \citet{Ghosh79} under
the assumption that the observed peak luminosity of $\sim$ 10\sS{36}
\lcgs is due to the accretion, 
the only possible solution for a NS with a mass of $M$
= 1.4 \Ms and a radius of $R$ = 10\sS{6} cm is with $\dot{P}$ $\sim$
$-$0.48 s yr\sS{-1} and a magnetic field of $B$ $\sim$
10\sS{13} G.  
The matching accretion rate
is $\dot{M}$ $\sim$ \nep{5}{15} g s\sS{-1} or \nep{8}{-11} \Ms yr\sS{-1}.  The
observed luminosity and thus the accretion rate are a bit low for
the estimated spin-up rate even with a strong $B$ field especially
considering that the X-ray emission and thus the spin-up trend have
likely been episodic.
If we use the historical average of
X-ray luminosities measured by \rxte ($\sim$ \nep{3}{36} \lcgs, K14), a
solution for the model by \citet{Ghosh79} is available for
the full range of $-\dot{P}$ $\sim$ 0.56 -- 1.7 s yr\sS{-1} with $B$ $\sim$
0.03 -- \nep{1.5}{12} G and $\dot{M}$ $\sim$ \nep{5.1}{-10} \Ms yr\sS{-1}.
With the duty cycle of $\sim$~30\%, 
the actual spin-up rate during the accretion has to be about three
times higher, which effectively means $-\dot{P}$ $\sim$ 1.9 -- 5.7 s
yr\sS{-1} during the accretion. 
The solution for the model by \citet{Ghosh79} exists for
a partial range of $-\dot{P}$ $\sim$ 1.9 -- 4.8 s yr\sS{-1} with $B$ $\sim$
0.44 -- \nep{29}{12} G and $\dot{M}$ $\sim$ \nep{8.4}{-10} \Ms yr\sS{-1}
during the accretion (or \nep{2.5}{-10} \Ms yr\sS{-1} on average).


The anti-correlation between the observed flux and the absorption
during pulsation as shown in Fig.~\ref{f:p}{\it f} 
suggests that the pulsation might be modulated with the periodic
occultation of the emission region by an
intervening absorber \cite[e.g.,][]{Suchy08}.  The modulated
absorption, however, cannot explain the missing soft X-ray component in
the early time of the observation when the strong pulsation was present
(Figs.~\ref{f:p}{\it g} and \ref{f:p}{\it h}).  
This implies that in the beginning the pulsar must have been inside or
behind a heavy absorber independent of the pulsation geometry.
A natural candidate for such an absorber is the circumstellar disk of the
companion.  The soft X-ray emission at the later time, then, indicates
that the NS came out of the circumstellar disk into view without
obscuration by the disk.  \citet{Coe15} showed that a similar picture
can explain the orbital eclipse in the long term X-ray lightcurve of
SXP5.05.  This analysis demonstrates for the first time the changes in
the X-ray emission as the NS crosses the circumstellar disk.  

\begin{figure} \begin{center}
\includegraphics*[width=0.450\textwidth,clip=true, trim=2cm 2cm 5cm 3cm] 
{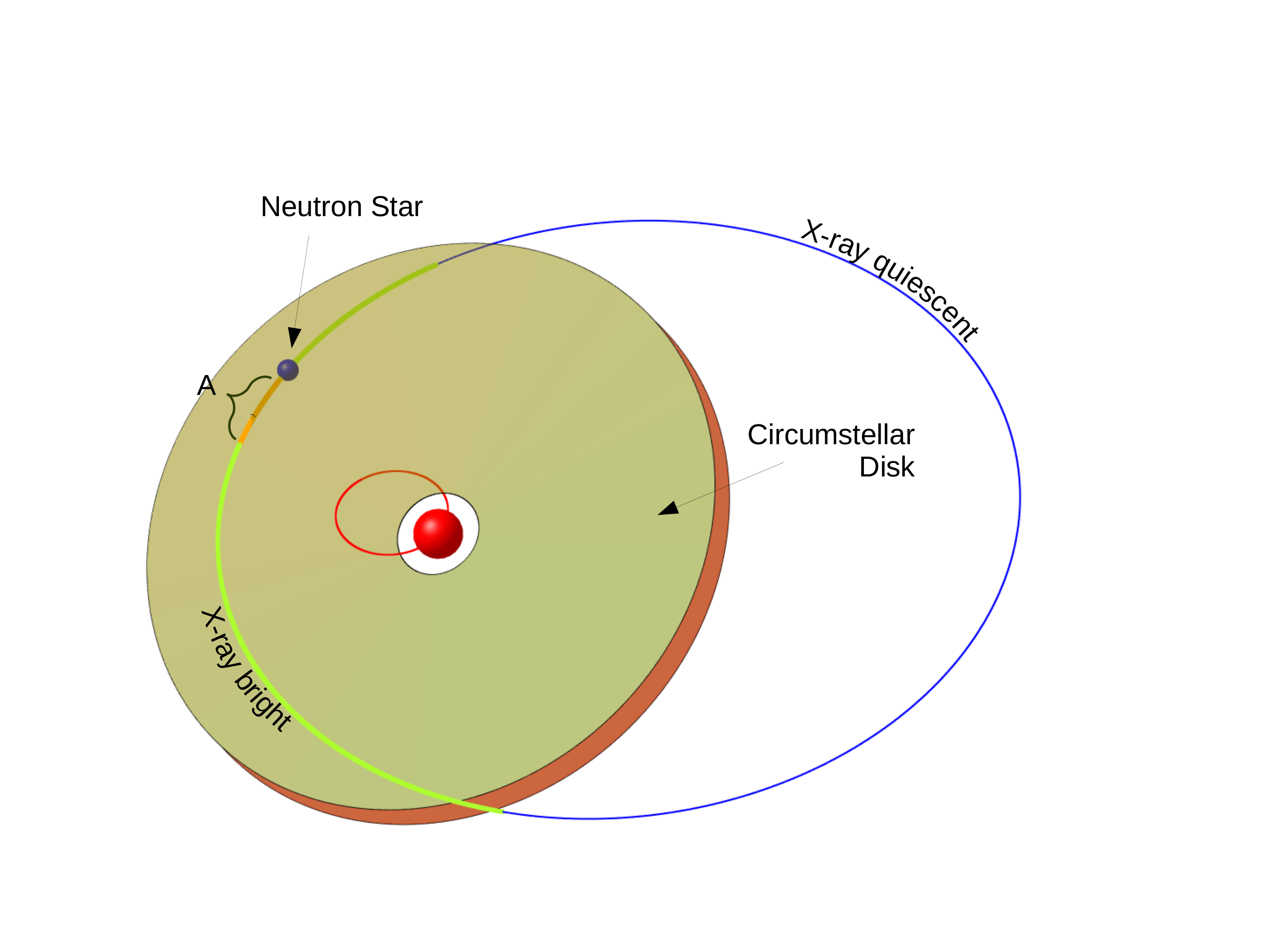} 
\caption{A possible orbital configuration of SXP214
under the assumptions of $M_1$ $\sim$ 10 \Ms, $M_2$ $\sim$ 1.4 \Ms,
$P_o$ $\sim$ 30 d, and $\delta$ $\sim$ 15\Deg. 
For about 25\% of the orbital period when the
NS is X-ray bright is marked with the thicker (green) line.
The interval marked 'A' (orange line) 
represents a possible observing period for Obs.~ID
14670. The red ellipse represents the orbit of the primary star.
}
\label{f:c}
\end{center}
\end{figure}

Alternatively one can consider accretion wake or clumpy winds 
as an absorber.  The stellar wind from a typical
B2-B3 star, however, is not strong enough to trigger accretion wake or
to generate clumpy winds.  For instance, clumpy winds in 4U~1700--37,
which cause an absorption of $\sim$ 10\sS{24} cm\sS{-1}, are from an
Of star with $\dot{M}$ $\sim$ 10\sS{-5} -- 10\sS{-6} \Ms yr\sS{-1}
\citep{Haberl89}, whereas the mass loss ($\dot{M}$) for B0-B1 stars
is already $\lesssim$ 10\sS{-9.5} \Ms yr\sS{-1} \citep{Smith14}.

In fact, we have likely witnessed the NS crossing
{\it through} (instead of behind) the circumstellar disk.
The observed spin-up trend and associated torque reversal favor the
accretion through a disk instead of quasi-spherical winds.
The constant intrinsic X-ray luminosity in Table~\ref{t:spec} indicates
that the accretion was in a steady state during our observation. This
appears to favor the NS crossing {\it behind} the circumstellar disk
without contact during our observation. But then it implies a relatively
long lifetime of the accretion disk and
a sharply truncated circumstellar disk for the NS to 
emerge in a short time ($\lesssim$ 30 ks).
Feeding enough material for the lasting accretion disk would likely
require a circumstellar disk with a high density, which in turn
would lead to a high peak X-ray luminosity ($>$10\sS{37} \lcgs; e.g.,
SXP5.05) and a long emission duty cycle, at odds with the long-term 
properties of SXP214.  

On the contrary, the observed rare episodic X-ray emission and 
the likely long orbital period suggest a relatively short lifetime of
the accretion disk and an only mildly perturbed circumstellar disk. A
relatively low X-ray luminosity also implies that the NS crosses
a low density region of the circumstellar disk (see below). The
relatively low optical emission (Fig.~\ref{f:o}{\it a}) during our
observation also implies a small circumstellar disk.  Thus, it is
plausible that the accretion disk became stable while feeding through the
circumstellar disk, and the intrinsic X-ray luminosity remained constant
for a while after the NS just emerged from the circumstellar disk.
Therefore, the NS crossing through the circumstellar disk provides a more
consistent picture.

Fig.~\ref{f:c} illustrates a possible orbital configuration of SXP214
under the assumptions of $M_1$ $\sim$ 10 \Ms, $M_2$ $\sim$ 1.4 \Ms, $P_o$
$\sim$ 30 d. The radius of the circumstellar disk shown in the figure
is about $12$ $R_*$ $\sim$ $60$ \Rs, and an inclination angle ($\delta$)
between the disk and the orbital plane is about 15\Deg. In this picture,
the NS crosses the circumstellar disk near the periastron passage once
per orbit and for about 25\% of the orbital period (depending on the
orbital motion induced change in the disk: e.g., warping), the first
Lagrangian point of the NS is on contact with the circumstellar disk,
during which the NS is potentially X-ray bright (thicker green line).
The interval marked by `A' (orange line) can be the 50 ks observing
period of Obs.~ID 14670, where the NS emerges from the disk in about
20--30 ks into the observation.

While the above illustration provides a possible picture consistent with
the observation including the X-ray emission duty cycle, note that how the
accretion phase remains relatively short is not clearly understood. In
particular, a caveat in the accretion disk for Type-I burst systems is
that the viscous time scale of a standard thin $\alpha$ disk is expected
to be longer than the orbital period, which should produce somewhat
persistent X-ray emission. \citet{Okazaki13} invokes a radiatively
inefficient accretion flow to explain relatively short X-ray emission
duty cycle for Type-I burst systems, where the accretion flow is X-ray
faint and the observed X-ray is mainly from the radiation at the poles.
Relatively high modulation amplitude ($\gtrsim$ 90\%) indeed suggests
that the observed X-ray emission from SXP214 is from the poles, but
regarding the exact accretion flow around the NS remains uncertain.

In our observation the soft X-ray component was hidden for at least 20 ks. 
Assuming the NS was already inside of the circumstellar disk in the
beginning of the observation based on 
the constant intrinsic luminosity and the orbital path is locally
straight, we can place a limit on the height
($H$) of the circumstellar disk using the travel distance of the NS in 20 ks.
For an inclination angle ($\delta$) between the disk and the orbital
plane,
\begin{eqnarray*}
	 H & \gtrsim &  (3.1 - 5.3) \times 10^{6}\ (M_{11.4}/P_{30})^{1/3}\ \sin(\delta)\  \mbox{km}. 
\end{eqnarray*}
If the majority of the observed extinction ($\sim$10\sS{24} cm\sS{-2})
at the early times is due to a simple slab-shape disk of a uniform 
density,\footnote{We use a simple model, given many unknown system
parameters.  Along with a possible density profile in the
circumstellar disk, the
orbital motion of the primary star (Fig.~\ref{f:c}) can cause warping in
the disk, generating additional density variation.}
the density ($\rho$) of the circumstellar disk is
\begin{eqnarray*}
	 \rho & \lesssim & (3.1 - 5.4) \times 10^{-12}\  (P_{30}/M_{11.4})^{1/3}
\sin(\delta)^{-1}  \sin(\theta)\ \mbox{g cm\sS{-3}},
\end{eqnarray*}
where $\theta$ is the angle between
the disk and the line of sight (i.e., $\theta$ = 0 for an edge-on view). 
The density of the circumstellar disk of Be stars
lies in the range between about 10\sS{-12} to a few times
10\sS{-10} g cm\sS{-3} \citep{Rivinius13}
and the disks of Be-XRBs are about 1.5 times denser \citep{Reig16}, 
so the above estimate for SXP214 indicates that
the section the NS crosses is of relatively low density.

The mass the NS accretes can be estimated using the volume swept by
the NS under the Bondi accretion.
\begin{eqnarray*}
	\dot{M} \sim 4 \pi \eta \rho \frac{G^2 M_2^2}{v_r^3} \frac{H}{v},
\end{eqnarray*}
where $v_r$ is the orbital speed of the NS relative to the gas in the
circumstellar disk, and  $\eta$ is the accretion fraction relative to the
circumstellar disk mass the NS disrupts as it crosses. 
Assuming the velocity of the gas in the circumstellar disk is in the
same order of the NS' orbital speed $v$, $v_r$ $\sim$ $\sqrt{2}$ $v$.
Therefore,
\begin{eqnarray*}
	\dot{M} & \sim &  (4.5 - 50) \times 10^{23} M_{1.4}^2 (P_{30}/M_{11.4})^{4/3} \sin(\theta) \mbox{\ \  g per passage} \\
		& = & (3.1 - 28) \times 10^{-9} M_{1.4}^2 P_{30}^{1/3}/M_{11.4}^{4/3} \sin(\theta) \mbox{\ \Ms yr\sS{-1}}. 
\end{eqnarray*}
If the NS captures about 1--10\% of the mass it disrupts 
(i.e., $\eta$ $\sim$ 0.01 -- 0.1),
the accretion rate meets what is required for the observed spin-up
process ($\sim$ \nep{2.5}{-10} \Ms yr\sS{-1}).


The spectral variation during the pulsation can be
understood by the periodic occultation of the emission region due to
the NS rather than a separate absorber.  
It is because the phase-resolved spectral
analysis of each interval in Fig.~\ref{f:p}{\it i}
shows that the local absorption is not likely the dominant factor for pulsation,
and the apparent correlation between 
the absorption and the pulse phase seen in the
full interval (Figs.~\ref{f:p}{\it c} and \ref{f:p}{\it f}) is likely
the results of the temporal evolution.  
For instance, the trailing
hard X-ray component during the ingress to
the off-phase can be due to Comptonized reflection from the
accretion stream. Alternatively, the soft X-rays may be mainly from a halo
around the bottom of the accretion column at the NS surface
\citep[e.g.,][]{Davidson73,
Lhyubarskii88, Kraus03}, which can explain the
total absence of the soft component during the off-phases 
in the pulse profile (red in Fig.~\ref{f:p}{\it e}).
A circular emission spot on the surface of a pulsar can produce the
observed eclipse-like feature in the pulse profile when combined with
general relativistic effects \citep[e.g.,][]{Wang81}.  

The remarkable variation in the X-ray emission from SXP214
indicates that the NS was caught while crossing the
circumstellar disk of the companion Be star.  X-ray observations provide
a rare opportunity to study the properties of the circumstellar disk
and the emission geometry of the NS surface.

\section{Acknowledgement}

VA acknowledges financial support from NASA/Chandra grants
GO3-14051X, AR4-15003X, NNX15AR30G and NASA/ADAP grant NNX10AH47G.
AZ acknowledges financial support from NASA/ADAP grant NNX12AN05G
and funding from the European Research Council under the European
Union's Seventh Framework Programme (FP/2007-2013)/ERC Grant Agreement n.~617001. 
JD, PP, and TG acknowledge financial support from NASA contract NAS8-03060.
MS acknowledges support by the the Deutsche Forschungsgemeinschaft
through the Heisenberg Programme (SA 2131/3-1). SL acknowledges
financial support from NASA/ADAP grant NNX14-AF77G.
The OGLE project has received funding from the National Science Centre,
Poland, grant MAESTRO 2014/14/A/ST9/00121 to AU.
We thank the anonymous referee for helpful comments and suggestions.

\end{document}